\documentclass[conference]{IEEEtran}
\usepackage{amsmath, amssymb, bm, cite, epsfig, psfrag}
\usepackage{graphicx}
\usepackage{soul} 
\usepackage[margin= 0.53 in]{geometry}
\usepackage{dblfloatfix}
\usepackage{array}
\usepackage{lipsum}

\usepackage[font=small]{caption}
\usepackage{epstopdf}	
\usepackage{longtable}
\usepackage{supertabular,booktabs}
\usepackage{bbm}
\usepackage{multirow}
\usepackage[usenames,dvipsnames]{xcolor}

\usepackage{subfigure}
\usepackage{etoolbox}
\usepackage{pbox}
\usepackage{fixltx2e}
\usepackage{tabu}	
\usepackage{filecontents}
\usepackage{enumerate}
\usepackage{textcomp}
\usepackage{colortbl}
\usepackage{fancyhdr}
\usepackage{bm}

\newtoggle{conference}

\togglefalse{conference} 
\graphicspath{{figures/}}

\newcolumntype{?}{!{\vrule width 2pt}}

\newcolumntype{P}[1]{>{\centering\hspace{0pt}}p{#1}}
\newcolumntype{M}[1]{>{\centering\hspace{0pt}}m{#1}}
\newcolumntype{L}{>{\centering\arraybackslash}m{3cm}}

\fancypagestyle{firststyle}{
	\fancyhf{}
	\fancyhead[L]{ \vspace{+0.6 cm} \small Y. Xing, T. S. Rappaport, and A. Ghosh, ``Millimeter Wave and sub-THz Indoor Radio Propagation Channel Measurements, Models, and Comparisons in an Office Environment (Invited Paper),'' \textit{IEEE Communications Letters}, June 2021, pp. 1-5.}     %% C or L or R.
}
\IEEEoverridecommandlockouts

\begin{document}
\title{Millimeter Wave and sub-THz Indoor Radio Propagation Channel Measurements, Models, and Comparisons in an Office Environment}\vspace{-1.0 cm}
\author{\IEEEauthorblockN{Yunchou Xing\IEEEauthorrefmark{1}, Theodore S. Rappaport\IEEEauthorrefmark{1}, and Amitava Ghosh\IEEEauthorrefmark{2}      (Invited paper)}%\vspace{-0.7cm}

\IEEEauthorblockA{	\small \IEEEauthorrefmark{1}NYU WIRELESS, NYU Tandon School of Engineering, Brooklyn, NY, 11201, \{ychou,  tsr\}@nyu.edu\\
	 \IEEEauthorrefmark{2}NOKIA Bell Labs, Naperville, IL, 60563, amitava.ghosh@nokia-bell-labs.com }\vspace{-1.0 cm}
					\thanks{This research is supported by the NYU WIRELESS Industrial Affiliates Program and NSF Research Grants: 1909206 and 2037845.}
}

\maketitle
\thispagestyle{firststyle}

\begin{abstract}

This letter provides a comparison of indoor radio propagation measurements and corresponding channel statistics at 28, 73, and 140 GHz, based on extensive measurements from 2014-2020 in an indoor office environment. Side-by-side comparisons of propagation characteristics (e.g., large-scale path loss and multipath time dispersion) across a wide range of frequencies from the low millimeter wave band of 28 GHz to the sub-THz band of 140 GHz illustrate the key similarities and differences in indoor wireless channels. The measurements and models show remarkably similar path loss exponents over frequencies \textcolor{black}{in both line-of-sight (LOS) and non-LOS (NLOS) scenarios}, when using a one meter free space reference distance, while the multipath time dispersion becomes smaller at higher frequencies. \textcolor{black}{The 3GPP indoor channel model overestimates the large-scale path loss and has unrealistic large numbers of clusters and multipath components per cluster compared to the measured channel statistics in this letter.}

%The empirical investigations used a wideband sliding correlation-based channel sounder with steerable horn antennas at both the transmitter and receiver to emulate futuristic electronic beam steering at both the mobile device and the access point.
% mention angular dispersion 
\end{abstract}
    

\section{Introduction}~\label{sec:intro}
The use of wide bandwidths (e.g., $\geq$100 MHz) in 5G and future wireless communication systems will enable multi-Gbps data rates for mobile devices and will usher in many new applications such as wireless cognition and centimeter-level positioning \cite{xing21a,rappaport19access,Kanhere20a}. Highly directional electronically-steered antenna arrays will be used by both handset terminals and base stations, resulting in directional wireless channels at mmWave, a significant departure from sub-6 GHz frequencies that use less directional antennas but undergo less penetration and diffraction loss from obstacles \cite{xing21a,rappaport19access,rappaport2013millimeter}. As shown in our companion paper \cite{xing21a}, the FCC has recently opened up spectrum above 95 GHz with four new unlicensed bands from 116 GHz to 246 GHz, ushering in a new era of wireless networks that will have hundreds of Gbps of throughput.

Propagation channels at mmWave frequencies in the 24-73 GHz range are somewhat different from sub-6 GHz channels in transmission properties yet are viable through the use of directional antennas on both ends of the link \cite{rappaport2013millimeter,Haneda16a,Ju20a,Sun14b,xing19GC,Sun16b}. However, there is very little known about indoor channels above 100 GHz that have a coverage range of 30-40 m. Furthermore, it is currently unknown whether sub-THz wireless networks (i.e., 100-300 GHz) have similar or contrasting propagation characteristics compared to radio channels at lower frequencies. Knowledge of channel characteristics at frequencies above 100 GHz, as well as key differences from lower frequencies, is vital for the creation of frequency-dependent channel models that can be applied over vast frequency ranges (e.g., from 1-300 GHz) to support the design of multi-band indoor wireless modems in global standard bodies such as 3GPP and IEEE \cite{3GPP2019}. Also, knowledge of how channel characteristics vary over wide frequency ranges can be useful for futuristic applications such as intelligent reflecting surfaces \cite{wu20IRS} and precise position location \cite{Kanhere20a}. Therefore, this work provides needed insights on indoor wireless channels from mmWave to sub-THz frequencies for futuristic wireless system designs. 

Previous channel measurements in frequency bands from 28 to 380 GHz \cite{pometcu18EuCAP,jacob09EUCAP,guan19TVT} have mostly focused on very close ranges (less than 10 m) due to the difficulty in achieving sufficient transmit power and large measurable path loss range \cite{rappaport19access}. Prior measurements at frequencies above 100 GHz \cite{ma18channel,abbasi20ICC,nguyen2018comparing} focused on line-of-sight (LOS) propagation using either reflective materials \cite{ma18channel} or an RF-over-fiber extension \cite{abbasi20ICC,nguyen2018comparing} of a VNA based system achieving over 100 m distance. However, no prior work has performed extensive indoor measurements wherein the same environment is used over a vast frequency range with commercially relevant distances of many tens of meters to provide comparisons of channel characteristics. 

\textcolor{black}{This paper presents the first comparison of the radio propagation characteristics from mmWave to sub-THz frequencies in an identical indoor office setting (the NYU WIRELESS center) based three different indoor propagation measurement campaigns at 28, 73, and 142 GHz with coverage distances up to 40 m.} Section \ref{sec:140PL} shows both indoor directional and omnidirectional large-scale path loss models at 142 GHz and compares the data to 28 and 73 GHz using identical locations in the same office. Section \ref{sec:Cstatistics} provides indoor channel statistics including delay spread, and the number of clusters and multipath components (MPCs) at 28, 73, and 142 GHz. Finally, concluding remarks in Section \ref{conclusion} show that, after normalizing over antenna gains and using a 1 meter free space reference distance, how remarkably similar the path loss exponents are for both LOS and non-LOS (NLOS) channels over 28-142 GHz. The multipath time dispersion becomes less prominent at higher frequencies meaning that it may be more difficult to exploit multipath diversity at THz frequencies without additive reflecting surfaces.

\section{Indoor Radio Propagation Measurements and Path Loss models at 28, 73, and 142 GHz}\label{sec:140PL}

Above 100 GHz, high phase noise and Doppler spread, limited output power due to device limitations, and the need to fabricate more compact directional phased arrays present challenges for the deployment of future wireless networks above 100 GHz \cite{rappaport19access,ma18channel,viswanathan20A, ghosh195g}. To overcome these challenges, it is critical that designers first understand the channel characteristics for radio frequencies above 100 GHz. 

\subsection{Measurement Systems and Environments}
To evaluate the similarity and differences of incumbent 5G bands with futuristic sub-THz bands above 100 GHz, extensive radio propagation measurements were conducted in an indoor hotspot (InH) office environment at 28, 73, and most recently 142 GHz over a period of 6 years since 2014 with TX-to-RX (TR) separation distances up to 40 m \cite{Ju20a}. 

Sliding correlation-based wideband channel sounder systems with identical steerable horn antennas at both the TX and RX for each frequency were used during the measurements \cite{Mac17JSACb,rappaport2013millimeter,xing18GC}, as shown in Table \ref{tab:sounder}. The channel sounders provide power delay profiles (PDPs) with dual-directional angular information. Five TX locations and 20 RX locations were selected across the entire floor, with the same locations used at 28, 73, and 142 GHz, resulting in 20 TX-RX combinations for each measured band (8 LOS and 12 NLOS locations) ranging from 3.9 m to 40.0 m. The measured office environment and measurement procedures are described in \cite{Ju20a,Mac15b}.

\begin{table}[]\caption{Summary of channel sounder systems and antennas used in measurements at 28, 73, and 142 GHz \cite{rappaport2013millimeter,Mac17JSACb,xing18GC}. The overall received power does not change with the signal bandwidth but the RX noise power and noise figure change \cite{Rap02a}. Thus, higher gain (more directional) antennas were used at both the TX and RX to compensate for noise and greater path loss in the first meter at higher frequencies.}\label{tab:sounder}
	\centering
	\begin{tabular}{lllll}
		\hline
		\multicolumn{1}{|l|}{RF Freq.}   & \multicolumn{1}{l|}{ RF Null-to-Null} &  \multicolumn{1}{l|}{Antenna} & \multicolumn{1}{c|}{Ant. Gain} \\ 
		\multicolumn{1}{|c|}{(GHz)}   & \multicolumn{1}{c|}{ Bandwidth} &  \multicolumn{1}{c|}{HPBW} & \multicolumn{1}{c|}{(dBi)} \\ \hline
		\multicolumn{1}{|c|}{28  \cite{rappaport2013millimeter}}           & \multicolumn{1}{c|}{0.8 GHz }       & \multicolumn{1}{c|}{30\textdegree}           & \multicolumn{1}{c|}{15.0  }              \\ \hline
		\multicolumn{1}{|c|}{73  \cite{Mac17JSACb}}      & \multicolumn{1}{c|}{0.8 GHz}     & \multicolumn{1}{c|}{15\textdegree}           & \multicolumn{1}{c|}{20.0 }              \\ \hline
		\multicolumn{1}{|c|}{142  \cite{xing18GC}}         & \multicolumn{1}{c|}{1.0 GHz}      & \multicolumn{1}{c|}{8\textdegree}            & \multicolumn{1}{c|}{27.0 }              \\ \hline
	\end{tabular}
\vspace{-0.5cm}
\end{table}

%\begin{figure}    
%	\centering
%	\includegraphics[width=0.350\textwidth]{140ChannelSounder2.png}
%	\caption{The 142 GHz channel sounder system at NYU, with TX antenna at 2.5 m above the floor close to the ceiling (2.7 m) to emulate indoor wireless access points and RX antenna positioned 1.5 m above the floor, similar to the height of a handset terminal.}
%	\label{fig:140CS}
%	\vspace{-0.3cm}
%\end{figure}

\subsection{Path Loss Models at 28, 73, and 142 GHz}
Radio propagation in mmWave and THz bands is range limited due to the severe path loss in the first meter and the limited output power of amplifiers, which are compensated by using directional antennas/arrays \cite{xing21a,rappaport19access}. \textcolor{black}{The measured path loss in this letter is calculated by:
\begin{equation}
\label{equ:PL}
\begin{split}
PL~\text{[dB]} &= P_t + G_t + G_r - P_r + G_{sym},
\end{split}
\end{equation} 
where $P_t$ is the output power fed into the TX antenna in dBm, $G_t$ and $G_r$ are TX and RX antennas gains in dBi, respectively, $P_r$ is the measured received power in dBm, and $G_{sym}$ is the processing gain of the channel sounder system in dB.}

The 1 m close-in (CI) free space reference distance model \eqref{equ:CI} is one of the most commonly used large-scale path loss models to predict the signal strength over distance, and is valid over different frequency bands \cite {Mac15b,rappaport2015wideband, sun2015path,3GPP2019,Ju20a}:
\begin{equation}
\label{equ:CI}
\small
\begin{split}
PL^{CI}(f_c,d_{\text{3D}})\ &= \text{FSPL}(f_c, \text{1 m}) +10n\log_{10}\left( \dfrac{d_{3D}}{\text{1 m}} \right)+ \chi_{\sigma},\\
\text{FSPL}(f_c,\text{1 m}) &= 32.4 + 20\log_{10}(\dfrac{f_c}{1\;\text{GHz}}),
\end{split}
\end{equation}
where \textcolor{black}{ $d_{\text{3D}}$ is the 3D separation distance between the TX and RX antennas,} FSPL$(f_c, 1 \;\text{m})$ is the free space path loss at carrier frequency $f_c$ in GHz at 1 m in dB, $n$ is the path loss exponent (PLE), and $\chi_{\sigma}$ is the shadow fading in dB (a zero mean Gaussian random variable with a standard deviation $\sigma$ in dB) \cite{rappaport2015wideband,Sun16b, sun2015path,Rap02a,Rap15a}.

\begin{table*}[!ht]
	\centering
	\caption{Directional InH-Office channel parameters of CI and CIF path loss models, RMS delay spread, the number of clusters, and the number of MPCs per cluster in both LOS and NLOS environments at 28, 73, and 142 GHz \cite{Mac15b,rappaport2015wideband,3GPP2019,Sun16b,Samimi15b,Ju20a}.}~\label{tab:PLcomp}
	\begin{tabular}{?c|c?c|c|c?c|c|c?c|c|c?}
		\hline
		\multicolumn{2}{?c?}{\textbf{Environment type}}                              & \multicolumn{3}{c?}{\textbf{LOS}} & \multicolumn{3}{c?}{$\textbf{NLOS}_{\textbf{Best}}$} & \multicolumn{3}{c?}{\textbf{NLOS}} \\ \hline
		\multicolumn{2}{?c?}{\textbf{Frequency [GHz]}}                       & \textbf{28 }   & \textbf{73  }& \textbf{142 }   & \textbf{28 } & \textbf{73 }  & \textbf{142}  & \textbf{28 }  & \textbf{73}   & \textbf{142} \\ \hline
		\multicolumn{2}{?c?}{\textbf{Directional Antenna HPBW}}                       & 30\textdegree   & 15\textdegree & 8\textdegree   & 30\textdegree  & 15\textdegree  & 8\textdegree  & 30\textdegree   & 15\textdegree   & 8\textdegree \\ \hline
		
		\multirow{1}{*}{\textbf{{Single-Band}}}      & n                        & 1.90  & 1.63       & 2.05   & 2.75  & 3.30       & 3.21   & 4.39   & 5.51   & 4.60       \\ \cline{2-11} 
		\multirow{1}{*}{\textbf{{Path Loss CI \cite{Sun16b,Samimi15b,Mac15b}}}} & $\sigma^{CI}$ {[}dB{]} & 3.38  & 3.06     & 2.89  & 7.00  & 8.76     & 6.03  & 7.30   & 8.94   & 13.80      \\ \hline \hline
		
		\multirow{1}{*}{\textbf{Multi-Band}}      & $n$    &\multicolumn{3}{c?}{1.86} & \multicolumn{3}{c?}{3.07}  & \multicolumn{3}{c?}{5.02} \\ \cline{2-11} 
		\multirow{1}{*}{\textbf{Path Loss CI \cite{Sun16b,Samimi15b,Mac15b} }}   & $\sigma^{CI}$ {[}dB{]} & \multicolumn{3}{c?}{3.45 }  & \multicolumn{3}{c?}{7.67}& \multicolumn{3}{c?}{13.97} \\ \hline \hline
		
		\multirow{1}{*}{\textbf{Multi-Band}}      & $n, b$    &\multicolumn{3}{c?}{$n=1.86, b=0.07$} & \multicolumn{3}{c?}{$n=3.07, b=0.05$}  & \multicolumn{3}{c?}{$n=5.02, b=0.03$} \\ \cline{2-11} 
		\multirow{1}{*}{\textbf{Path Loss CIF \cite{Sun16b,Samimi15b,Mac15b} }}   & $\sigma^{CIF}$ {[}dB{]} & \multicolumn{3}{c?}{3.45 }  & \multicolumn{3}{c?}{7.67}& \multicolumn{3}{c?}{13.85} \\ \hline \hline
		
		\multirow{3}{*}{\textbf{{RMS Delay Spread} \cite{Mac15b,rappaport2015wideband}}} 
		& $\min_{DS}$   [ns]            & 0.87    & 0.76         & 0.69     & 0.92    & 3.74          & 0.60      & 0.57      & 0.51   & 0.28\\
		\cline{2-11} 
		& $\max_{DS}$   [ns]            & 5.50     & 5.34        & 11.94      & 44.49     & 31.37          & 10.76      & 198.55      & 141.97   & 92.45\\ \cline{2-11} 
		& $\mu_{DS}$ [ns]               & 3.85    & 3.53          & 2.71      & 10.23     & 7.39          & 5.65      & 17.64      & 12.50   & 8.86          \\ \cline{2-11} 
%		& $\sigma_{DS}$   [ns]            & 1.64     & 1.61         & 1.88      & 11.09     & 6.08          & 5.38      & 14.92      & 15.15   & 9.95 \\ 
		\hline \hline
		% Table 6 in indoor15ACCESS

		%%%%%%%%%% NUM OF CLUSTER needs double check
		\multirow{2}{*}{{\textbf{NumCluster \cite{Ju20a}}}}       & $\mu_{NC}$               & 1.41     & 1.32          & 1.25       & 1.65     & 1.48          &  1.16     & 3.41      & 2.60   & 2.39          \\ \cline{2-11} 
		& $\sigma_{NC}$            & 0.85    & 0.96         & 0.94      & 0.78     & 0.89          & 0.69      & 1.96      & 1.70   & 1.48          \\ \hline \hline
		
		\multirow{2}{*}{{\textbf{NumMPCperCluster \cite{Ju20a}}}} & $\mu_{MPC}$              & 2.45     & 2.53          & 2.11      & 2.56     & 2.44          & 1.98      & 3.16      & 2.80   & 1.18          \\ \cline{2-11} 
		& $\sigma_{MPC}$           & 2.19     & 2.27          & 1.43      & 1.54     & 2.18          & 2.26      & 4.56      & 5.20   & 2.21          \\ \hline 
	\end{tabular}
\vspace{-0.5cm}
\end{table*}

The indoor directional LOS PLE for the best-fit CI path loss model is $n=$ 1.90 at 28 GHz, $n=$ 1.63 at 73 GHz, and is $n=$ 2.05 at 142 GHz with a shadow fading standard deviation of $\sigma=$ 3.38 dB, 3.06 dB, and 2.89 dB, respectively, as presented in Table \ref{tab:PLcomp}. These values all are close to the theoretical free space PLE value of 2.0, with lower mmWave frequencies experiencing waveguiding effects and a PLE less than 2.0. This hints at the fact that higher sub-THz frequencies have fewer reflections in LOS directional channels (e.g., less waveguiding) due to narrower beam antennas at both ends of the link that attenuate energy toward the reflecting surfaces of walls, ceiling, and floor (as shown in Table \ref{tab:sounder}). The LOS measurements show that there is 2-4 dB more average loss at 10 m and 3-7 dB more average loss at 40 m at 142 GHz compared to the average loss at 28 and 73 GHz when referenced to a one meter free space reference distance \cite{Sun16b,rappaport19access,Sun14a}.

The $\text{NLOS}_{\text{Best}}$ is defined as the best pointing direction of both the TX and RX antennas for which the maximum power is received at the RX at each of the NLOS measurement locations. The best-fit $\text{NLOS}_{\text{Best}}$ PLEs are $n=$ 2.75, 3.30, and 3.21 at 28, 73, and 142 GHz, respectively, with shadow fading standard deviations of $\sigma=$ 7.00, 8.76, and 6.03 dB, respectively, showing that the indoor $\text{NLOS}_{\text{Best}}$ channels at 73 and 142 GHz are more lossy (higher PLEs) than the channels at 28 GHz. 

Two transmission properties, reflections and penetrations, appear to cause the slight difference of the NLOS PLEs at 28 and 142 GHz \cite{rappaport19access,xing19GC}, since it is known that as frequencies increase, there is less transmissivity through reflecting objects and more power is reflected (i.e., greater power in the reflection direction but less penetrated power is observed at 142 GHz compared to 28 and 73 GHz)\cite{xing19GC,ju19icc}. However, the surfaces are relatively rough at higher frequencies, and indoor reflection and scattering measurements at 28, 73, and 142 GHz show that the scattered power is negligible (more than 20 dB less) compared to the signal power in the reflection direction for most indoor materials (e.g., drywall and clear glass) \cite{xing19GC,ju19icc}.

Omnidirectional path loss models were developed from the directional measurements by synthesizing omnidirectional antenna patterns, received power, and path loss from directional antennas \cite{sun2015synthesizing,rappaport2015wideband}. The LOS omnidirectional PLE at 142 GHz is 1.74 with a shadow fading standard deviation of 3.62 dB, which, as expected, is less than the LOS directional PLE of 2.05 at the same 142 GHz due to the capture of energy over the entire horizon using an omnidirectional pattern \cite{sun2015synthesizing,rappaport2015wideband}. The LOS omnidirectional channel offers 3.1 dB less average loss at 10 m and 5.0 dB less average loss at 40 m than the LOS directional channel at 142 GHz (with antenna gains removed). However, in practice, omnidirectional antennas cover a shorter link range due to the lower antenna gain. The NLOS omnidirectional PLE at 142 GHz is 2.83 with a shadow fading standard deviation of 6.07 dB which is a much lossier channel than LOS, yet offers better link coverage than the $\text{NLOS}_{\text{Best}}$ PLE of 3.21 and the arbitrary directional NLOS PLE of 4.60 at 142 GHz as shown in Tables \ref{tab:PLcomp} and \ref{tab:Omnicomp}. The higher PLE in NLOS at higher frequencies suggest accurate beamforming algorithms will be needed to find, capture, and combine the most dominant multipath energy to maintain indoor NLOS communication links above 100 GHz \cite{rappaport19access,Sun14a,Sun14b}.

\begin{figure}    
	\centering
	\includegraphics[width=0.45\textwidth]{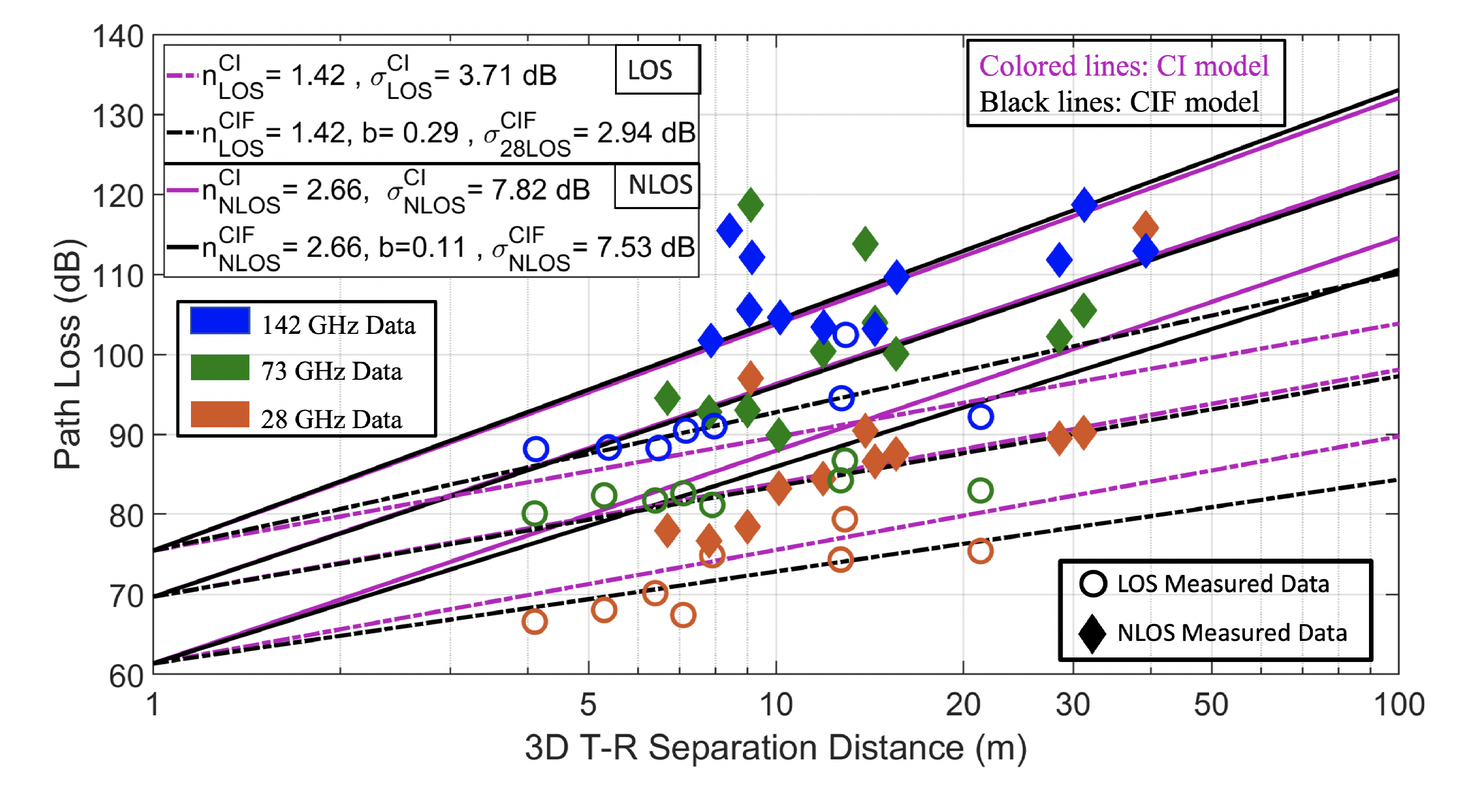}
	\caption{InH-Office 28, 73, and 142 GHz multi-band omnidirectional CI and CIF path loss models for both LOS and NLOS scenarios with antenna gains removed and with respect to a 1 m free space reference \cite{Sun16b,Mac15b,Samimi15b}. The diamonds and circles represent measured omnidirectional path loss in NLOS and LOS locations, respectively \cite{Ju20a,Mac15b,xing19GC}. The fixed reference frequency $f_0$ for the CIF path loss model is 81 GHz for both LOS and NLOS conditions. Results here and Table \ref{tab:Omnicomp} show the 142 GHz band is lossier in NLOS conditions, with the CI and CIF models being remarkably similar \cite{Sun16b,Mac15b}.}
	\label{fig:140PL}
	\vspace{-0.5cm}
\end{figure}

The omnidirectional PLE and shadowing parameters of the CI path loss model with 1 m free space reference distance for both LOS and NLOS over each of the three bands and for all bands combined are summarized in Table \ref{tab:Omnicomp}. The indoor omnidirectional LOS PLEs and shadow fading standard deviations $(n, \sigma)$ are (1.17, 2.72 dB), (1.36, 2.30 dB), and (1.74, 3.62 dB) at 28, 73, and 142 GHz bands, respectively, and the higher LOS PLE at 142 GHz indicates that the MPCs undergo penetration, reflection, and absorption with more loss at 142 GHz than at 28 and 73 GHz \cite{Rap15a,xing19GC}. In the NLOS omnidirectional case, the PLEs and shadow fading standard deviations $(n, \sigma)$ are  (2.37, 7.22 dB), (2.81, 8.71 dB), and (2.83, 6.07 dB) at 28, 73, and 142 GHz, respectively, showing that 28 GHz is less lossy than both 73 and 142 GHz, with the two higher bands behaving very similarly, except for the first meter loss. 

The CI model with a frequency-weighted PLE (CIF) path loss model \cite{Mac15b,Sun16b} was proposed as a viable multi-band model for indoor path loss and can be considered as an extension of the CI path loss model to offer an extra degree of freedom for more accurate statistical modeling over a wide range of frequencies:
\begin{equation}
\label{equ:CIF}
\small
\begin{split}
PL^{CIF}&(f_c,d_{\text{3D}})\;\text{[dB]} = \text{FSPL}(f_c, \text{1 m})+\\
  &10n\left( 1+ b \left( \dfrac{f-f_0}{f_0}\right) \right) \log_{10}\left( \dfrac{d}{\text{1 m}} \right) + \chi^{CIF}_{\sigma},
\end{split}
\end{equation}
where $n$ is the PLE at the weighted frequency average $f_0$ of all measurements for each specific environments, $b$ is a model fitting parameter that presents the slope of linear frequency dependency of path loss, and $n\left( 1+ b ( {f-f_0})/{f_0} \right)$ represents the frequency-dependent PLE at frequency $f$. The weighted average frequency $f_0$ is computed over $K$ frequency bands as $f_0=\sum_{k=1}^{K}f_kN_k/\sum_{k=1}^{K}N_k$, where $N_k$ is the number of measurements at a particular frequency $f$ \cite{Mac15b,Sun16b}. 

%The CIF path loss model \eqref{equ:CIF} uses two parameters to model average path loss over distance over multiple frequency bands, and reverts to the single-parameter CI model \eqref{equ:CI} when $b=0$ or when using data from just one frequency band \cite{Sun16b,Mac15b,Samimi15b}. The introduction and closed-form expressions for the best-fit model parameters of the CIF model can be found in \cite{Mac15b,Sun16b} and is expected to be useful over the wide swath for bands going up to THz \cite{xing21a,Mac15b}.

In this paper, we used the data to calculate $f_0 =$ 81 GHz for both LOS and NLOS scenarios. In LOS omnidirectional scenarios, $n^{CIF}_{LOS}=1.42$ and $b^{CIF}_{LOS}=0.29$ are the best-fit parameters of the omnidirectional CIF path loss model with a shadow fading standard deviation of $\sigma^{CIF}_{LOS} = 2.94$ dB, as shown in Table \ref{tab:Omnicomp}, when applied to all measurements across 28, 73, and 142 GHz. In NLOS omnidirectional scenarios, $n^{CIF}_{NLOS}=2.66$ and $b^{CIF}_{NLOS}=0.11$ are the best-fit parameters of the omnidirectional CIF path loss model with a shadow fading standard deviation of $\sigma^{CIF}_{NLOS} = 7.53$ dB.

%, as shown in Table \ref{tab:Omnicomp}.

Note that the CI path loss model \eqref{equ:CI} can also be used as a multi-band path loss model, using only a single parameter PLE \cite{Mac15b, Samimi15b, Sun16b}. Fig. \ref{fig:140PL} shows the indoor office 28, 73, and 142 GHz multi-band omnidirectional CI and CIF path loss models for both LOS and NLOS scenarios with antenna gains removed at both ends of the link and with respect to a 1 m free space reference. CI and CIF path loss models show remarkable similarity in terms of path loss exponents over 28, 73, and 142 GHz bands, when referenced to the first meter free space reference distance \cite{Sun16b,Mac15b,Samimi15b}, implying the extra parameter in the CIF model may not be necessary over wide frequency bands. This illuminates the fact that an identical PLE may accurately model path loss over a vast range of frequencies, with the only frequency-dependent effect being \textit{the path loss in the first meter of propagation as energy spreads into the far field} \cite{xing21a,Mac15b, Samimi15b, Sun16b}. 

%, since the large-scale path loss indoor THz channels are very similar modeled with either the CI or CIF model

%Directional and omnidirectional multi-band CI and CIF path loss models are summarized in Table \ref{tab:PLcomp} and \ref{tab:Omnicomp} for both LOS and NLOS scenarios, with both models providing nearly identical results with respect to the received power and shadow fading standard deviation.

\begin{table*}[]
		\centering
	\caption{Omnidirectional InH-Office channel parameters of CI and CIF path loss models, RMS delay spread, the number of clusters, and the number of MPCs per cluster at 28, 73, and 142 GHz compared to 3GPP InH-Office standard \cite{Mac15b,rappaport2015wideband,3GPP2019,Sun16b,Samimi15b,Ju20a}.}~\label{tab:Omnicomp}
	\begin{tabular}{?c|c?c|c|c|c|c?c|c|c|c|c?}
		\hline
		\multicolumn{2}{?c?}{\textbf{Environment type}}                       & \multicolumn{5}{c?}{\textbf{InH-Office LOS}} & \multicolumn{5}{c?}{\textbf{InH-Office NLOS}} \\ \hline
		
		\multicolumn{2}{?c?}{\textbf{omnidirectional NYU \& 3GPP}}                       & \multicolumn{3}{c|}{\textbf{NYU WIRELESS}} &\multicolumn{2}{c?}{\textbf{3GPP InH-Office}} &\multicolumn{3}{c|}{\textbf{NYU WIRELESS}} &\multicolumn{2}{c?}{\textbf{3GPP InH-Office}}\\ \hline
		\multicolumn{2}{?c?}{\textbf{Frequency [GHz]}}                    & \textbf{28}    &\textbf{ 73}      & \textbf{142}  & \textbf{28} & \textbf{73}    & \textbf{28}     & \textbf{73}      & \textbf{142}  &\textbf{28}&\textbf{73 }  \\ \hline
		
			\multirow{1}{*}{\textbf{Single-Band}}      & n    & 1.17   & 1.36   & 1.74 & N/A & N/A   & 2.37   & 2.81    & 2.83 & N/A  & N/A   \\ \cline{2-12} 
			\multirow{1}{*}{\textbf{Path Loss CI \cite{Sun16b,Samimi15b,Mac15b}}}  & $\sigma^{CI}$ {[}dB{]} & 2.72   & 2.30   & 3.62  & N/A& N/A& 7.22   & 8.71   & 6.07 & N/A & N/A \\ \hline \hline
			
		\multirow{1}{*}{\textbf{Multi-Band}}      & $n$    &\multicolumn{3}{c|}{$1.42$} & \multicolumn{2}{c?}{1.73}  & \multicolumn{3}{c|}{2.66}& \multicolumn{2}{c?}{3.19} \\ \cline{2-12} 
		\multirow{1}{*}{\textbf{Path Loss CI \cite{Sun16b,Samimi15b,Mac15b} }}   & $\sigma^{CI}$ {[}dB{]} & \multicolumn{3}{c|}{3.71 }  & \multicolumn{2}{c?}{3.00}& \multicolumn{3}{c|}{7.82}&\multicolumn{2}{c?}{8.29}    \\ \hline \hline
		
		\multirow{1}{*}{\textbf{Multi-Band}}      & $n,b$    &\multicolumn{3}{c|}{$n=1.42, b =0.29$} & N/A & N/A  & \multicolumn{3}{c|}{\textbf{$n=2.66, b =0.11$}}& N/A & N/A  \\ \cline{2-12} 
		\multirow{1}{*}{\textbf{Path Loss CIF \cite{Sun16b,Samimi15b,Mac15b}}}   & $\sigma^{CIF}$ {[}dB{]} & \multicolumn{3}{c|}{2.94 }  & N/A & N/A& \multicolumn{3}{c|}{7.53}&N/A   & N/A    \\ \hline \hline

		% minDS 140 needs to be double checked
		\multirow{3}{*}{\textbf{ RMS Delay Spread \cite{Mac15b,rappaport2015wideband}}} & $\min_{DS}$ [ns] & 0.70      & 0.60     & 0.71 & N/A & N/A    & 0.60      & 0.50   & 0.60   & N/A & N/A  \\ \cline{2-12} 
		& $\max_{DS}$  [ns]            & 134.40      & 101.90     &11.94   &N/A   & N/A  &198.50      & 142.00    & 60.87   & N/A & N/A \\ \cline{2-12} 
		& $\mu_{DS}$ [ns]                 & 10.80     & 6.24      & 3.00  & 20.40  & 20.21  & 17.10     & 12.30   & 9.20  &27.40 &21.52   \\ \cline{2-12} 
%		& $\sigma_{DS}$  [ns]            & 4.24     & 4.11    &2.14  & 5.68  & 5.62   & 13.80      & 14.2    & 8.74 & 8.57 &8.17    \\ 
\hline \hline

		\multirow{2}{*}{\textbf{NumCluster \cite{Ju20a}}}       & $\mu_{NC}$       & {4.60}     & {2.76}      & 1.90     &15&15  & {5.40}     &{ 3.20 }   & 2.80  &19 &19    \\ \cline{2-12} 
		& $\sigma_{NC}$            & {1.94}      & {2.32}     & 1.30    & N/A& N/A  & {1.96}      & {1.70}    & 1.65   & N/A& N/A   \\ \hline \hline
		
		\multirow{2}{*}{\textbf{NumMPCperCluster \cite{Ju20a}}} & $\mu_{MPC}$   & {4.70}     & {3.43}      & 2.40   &20 &20  & {6.40}     & {3.20}    & 2.20 &20 &20    \\ \cline{2-12}
		& $\sigma_{MPC}$           &  {3.65}    & {2.86}      & 2.20  &N/A  &N/A & {4.58}     & {5.20}    & 2.47 & N/A &N/A    \\ \hline 
		
	\end{tabular}
\vspace{-0.5cm}
\end{table*}

\section{Multipath Statistics at 28, 73, and 142 GHz}\label{sec:Cstatistics}
The indoor wireless channel statistics at 28, 73, and 142 GHz bands are summarized in Tables \ref{tab:PLcomp} and \ref{tab:Omnicomp} for directional and omnidirectional channels, respectively.  %The spatial statistical channel modeling approach introduced in \cite{Ju20a} is used in this letter. 

%The large-scale path loss, shadow fading, root mean square delay spread (RMS DS), the number of time clusters, as well as the number of MPCs per cluster were evaluated from the measured data.

\subsection{RMS Delay Spread} \label{sec:DS}
RMS Delay spread characterizes the multipath richness, time dispersion, and coherence bandwidth of a radio propagation channel, depending on the signal's bandwidth \cite{Rap02a,Rap15a}. A 5 dB SNR threshold relative to the mean thermal noise floor of a raw PDP was used for detecting and keeping the MPCs in each PDP \cite{rappaport2015wideband}. \textcolor{black}{The number of MPCs are computed by peak tracking algorithms as illustrated in \cite{Ju20a} that one peak in a PDP is counted as a valid MPC if the time delay between its next adjacent peak (above the 5 dB SNR threshold) is larger than the channel sounder time resolution (the inverse of the channel bandwidth).} 

%The minimum, maximum, and mean of RMS delay spread values from measurements in an identical indoor office setting at 28, 73, and 142 GHz are presented in Table \ref{tab:PLcomp} and \ref{tab:Omnicomp} for both directional and omnidirectional indoor channels, respectively, in both LOS and NLOS scenarios \cite{rappaport2015wideband}. 

As shown in Table \ref{tab:PLcomp}, there is negligible RMS DS in LOS directional scenarios with a mean of 3-4 ns, which is close to the width of the channel sounder's probing signal across these three frequency bands when directional antennas are used to line up antenna boresights (our probe had a 3 dB power width of 2.5 ns at 28 and 73 GHz and 2 ns at 142 GHz as shown in Table \ref{tab:sounder}). The excess delays of the multipath components reflected from the walls, ceiling, and floor would be within (and cause a spread of) the original pulse width.

%where both the TX and RX antennas are pointing in the direction with maximum received power at the RX, 

In $\text{NLOS}_{\text{Best}}$ scenarios, the minimum RMS DS ($\min_{DS}$) is extremely small (e.g., 0.5 ns) and similar to the minimum RMS DS in LOS directional environments across 28, 73, and 142 GHz bands, however, the mean RMS DS ($\mu_{DS}$) increases by a few ns (e.g., 3-6 ns) at all frequencies compared to LOS directional scenarios, and the maximum RMS DS ($\max_{DS}$) was observed to be 44 ns at 28 GHz, 31 ns at 73 GHz, and 11 ns at 142 GHz, with 90\% of the RMS DS less than 40 ns, 25 ns, and 9 ns at 28, 73, and 142 GHz bands, respectively. The data clearly show maximum observable RMS DS is frequency dependent, with much less maximum delay spread at higher frequencies.

%In NLOS directional scenarios where the directional antenna at the RX was arbitrarily pointed, the maximum delay spread was observed to be more than 90 ns with only 3\%, 2\%, and 1\% of the RMS DS larger than 90 ns at 28, 73, and 142 GHz bands \cite{Ju20a,Mac15b}, respectively. The mean RMS DS decreases as frequency increases with 18 ns at 28 GHz, 13 ns at 73 GHz, and 9 ns at 142 GHz, indicating that the time dispersion is smaller at higher frequencies, likely due to the narrower beam antennas used at higher frequencies that fail to capture all the multipath in the environment and miss reflected multipath from directions out of antenna pointing beam \cite{rappaport2015wideband,Ju20a,Mac15b}. 

The omnidirectional RMS DS obtained from measurements for both the LOS and NLOS scenarios at 28, 73, and 142 GHz bands \cite{Ju20a,Mac15b} are presented in Table \ref{tab:Omnicomp} and are compared with mean $\mu_{DS}$ and standard deviation $\sigma_{DS}$ values from 3GPP TR 38.901 Release 16 (see Table 7.5-6 in \cite{3GPP2019}). Note that 3GPP does not specify the RMS DS for channels above 100 GHz. 

In LOS omnidirectional scenarios, the mean RMS DS decreases with increasing frequency and is 11, 6, and 3 ns at 28, 73, and 142 GHz bands, respectively. By contrast, the mean RMS DS predicted by the 3GPP InH channel model is virtually identical at 28 and 73 GHz (20.40 ns and 20.21 ns, respectively), however, in our work the measured mean RMS DS at 73 GHz decreases by 40\% (about 5 ns lower) than the measured mean RMS DS at 28 GHz, and the mean RMS DS decreases by 50\% further to a very small 3 ns at 142 GHz.

In NLOS omnidirectional scenarios, when compared to LOS omnidirectional channels, the mean RMS DS increases by 60\%, 100\%, and 200\%  at 28, 73, and 142 GHz, respectively, with mean values of 17, 12, and 9 ns at 28, 73, and 142 GHz bands, respectively, indicating a wider time spread of multipath energy that is frequency dependent in NLOS environments. The free space path loss in the first meter and partition losses are larger at higher frequencies, such that the signal power of MPCs with large time delays (longer propagation distances and are more likely to be blocked, and hence have weaker power) may be below the noise floor and not detected by the channel sounder RX \cite{rappaport2015wideband}. This is one cause for why the RMS DS at 142 GHz is lower than the RMS DSs at 73 and 28 GHz. 

%In general, the RMS DS in NLOS locations is larger than LOS locations, since obstructions in NLOS locations block or severely attenuate the direct path, leading to larger RMS DSs \cite{rappaport2015wideband,Mac15b,Ju20a}.

\subsection{The Numbers of Time Clusters and MPCs per Cluster}
The numbers of time clusters and MPCs per cluster depends on the minimum inter-cluster time void interval (MTI), which is the minimum time interval between two separate time clusters \cite{Ju20a,samimi20163,rappaport2013millimeter}. There are fewer time clusters but more MPCs per cluster when the MTI is larger. An MTI of 6 ns \cite{Ju20a} (determined by the width of corridors, 2 m) is used in this letter, based on the idea that the physical environment helps delineate the observed temporal cluster partitions of multipath energy arriving at a receiver \cite{samimi20163,rappaport2013millimeter}. 

The number of time clusters follows a Poisson distribution at all three frequencies as shown in \cite{Ju20a}. The omnidirectional mean numbers of time clusters are found to decrease with increasing frequency and are 4.60, 2.76, and 1.90 at 28, 73, and 142 GHz, respectively, in indoor LOS scenarios, and are 5.40, 3.20, and 2.80 at 28, 73, and 142 GHz in indoor NLOS scenarios, again showing the frequency dependent nature of multipath. The channels are more sparse in both LOS and NLOS omnidirectional scenarios at higher frequencies due to the larger partition loss (e.g., 4-8 dB higher loss at 142 GHz than 28 GHz for different materials \cite{xing19GC}) which cause MPCs with longer propagation times and weaker power to fall below the noise floor. Using smart antennas to capture the most dominant multipath energy for beam combining to achieve range extension is needed as wireless moves into the THz regime \cite{rappaport19access,Sun14a,Samimi15b,Sun14b}. 

The number of MPCs per cluster follows a composite of a delta function $\delta (n-1)$ and a discrete exponential distribution at all three frequencies \cite{Ju20a}. At 28 GHz, the mean number of MPCs per cluster in the NLOS omnidirectional scenarios is larger than in the LOS omnidirectional scenarios. However, there are fewer MPCs per cluster in the NLOS scenarios than in the LOS scenarios at 73 and 142 GHz, since the larger partition loss in NLOS scenarios makes the channels more sparse at higher frequencies \cite{xing19GC,Ju20a}.

\section{Conclusion}\label{conclusion}
Indoor mmWave and sub-THz wireless channels at 28, 73, and 142 GHz are compared based on extensive radio propagation measurements in a typical indoor office environment. The indoor office large-scale path loss results (CI and CIF path loss models) show that there is remarkable similarity in terms of path loss exponents over 28, 73, and 142 GHz for both LOS and NLOS scenarios, when referenced to the first meter free-space reference distance \cite{Sun16b,Mac15b,rappaport2013millimeter}. The results imply that THz channels are similar to today's mmWave wireless propagation channels except for the path loss in the first meter of propagation when energy spreads into the far field. \textcolor{black}{Our results differ from existing 3GPP InH prediction models, and show strong frequency dependence on multipath time dispersion, with much less time dispersion at higher frequencies.} Mathematical distributions of the number of multipath clusters, RMS delay spread, the number of multipath components or subpaths per cluster can be applied for frequencies above and below 100 GHz, although the statistical means of those distributions decrease with increasing frequency. The measurements and models presented here will help in the development of frequency-dependent indoor office channel models that can be applied over vast mmWave and THz frequency ranges.

\bibliographystyle{IEEEtran}
\bibliography{Indoor140GHznew}

% Generated by IEEEtran.bst, version: 1.14 (2015/08/26)
\begin{thebibliography}{10}
\providecommand{\url}[1]{#1}
\csname url@samestyle\endcsname
\providecommand{\newblock}{\relax}
\providecommand{\bibinfo}[2]{#2}
\providecommand{\BIBentrySTDinterwordspacing}{\spaceskip=0pt\relax}
\providecommand{\BIBentryALTinterwordstretchfactor}{4}
\providecommand{\BIBentryALTinterwordspacing}{\spaceskip=\fontdimen2\font plus
\BIBentryALTinterwordstretchfactor\fontdimen3\font minus
  \fontdimen4\font\relax}
\providecommand{\BIBforeignlanguage}[2]{{%
\expandafter\ifx\csname l@#1\endcsname\relax
\typeout{** WARNING: IEEEtran.bst: No hyphenation pattern has been}%
\typeout{** loaded for the language `#1'. Using the pattern for}%
\typeout{** the default language instead.}%
\else
\language=\csname l@#1\endcsname
\fi
#2}}
\providecommand{\BIBdecl}{\relax}
\BIBdecl

\bibitem{xing21a}
Y.~Xing and T.~S. Rappaport, ``{Terahertz Wireless Communications: Co-sharing
  for Terrestrial and Satellite Systems above 100 GHz (Invited)},'' \emph{IEEE
  Communications Letters}, pp. 1--5, June 2021.

\bibitem{rappaport19access}
T.~S. Rappaport \emph{et~al.}, ``{Wireless Communications and Applications
  Above 100 GHz: Opportunities and Challenges for 6G and Beyond (Invited)},''
  \emph{IEEE Access}, vol.~7, pp. 78\,729--78\,757, Feb. 2019.

\bibitem{Kanhere20a}
O.~Kanhere and T.~S. Rappaport, ``{Position Location for Futuristic Wireless
  Communications: 5G and Beyond},'' \emph{IEEE Communications Magazine}, pp.
  70--75, Feb. 2021.

\bibitem{rappaport2013millimeter}
T.~S. Rappaport \emph{et~al.}, ``{Millimeter Wave Mobile Communications for
  {5G} Cellular: It Will Work!}'' \emph{IEEE Access}, vol.~1, pp. 335--349, May
  2013.

\bibitem{Haneda16a}
K.~Haneda \emph{et~al.}, ``{5G 3GPP}-like channel models for outdoor urban
  microcellular and macrocellular environments,'' in \emph{2016 IEEE 83rd
  Vehicular Technology Conference (VTC2016-Spring)}, May 2016, pp. 1--7.

\bibitem{Ju20a}
S.~Ju \emph{et~al.}, ``{Millimeter Wave and Sub-Terahertz Spatial Statistical
  Channel Model for an Indoor Office Building},'' \emph{IEEE Journal on
  Selected Areas in Communications}, April 2021.

\bibitem{Sun14b}
S.~Sun \emph{et~al.}, ``{MIMO} for millimeter-wave wireless communications:
  beamforming, spatial multiplexing, or both?'' \emph{IEEE Communications
  Magazine}, vol.~52, no.~12, pp. 110--121, Dec. 2014.

\bibitem{xing19GC}
Y.~Xing and T.~S. Rappaport, ``{Indoor Wireless Channel Properties at
  Millimeter Wave and Sub-Terahertz Frequencies},'' in \emph{\textit{IEEE 2019
  Global Communications Conference}}, Dec. 2019, pp. 1--6.

\bibitem{Sun16b}
S.~Sun \emph{et~al.}, ``{Investigation of Prediction Accuracy, Sensitivity, and
  Parameter Stability of Large-Scale Propagation Path Loss Models for {5G}
  Wireless Communications (Invited Paper)},'' \emph{IEEE Transactions on
  Vehicular Technology}, vol.~65, no.~5, pp. 2843--2860, May 2016.

\bibitem{3GPP2019}
3GPP, ``{Study on channel model for frequencies from 0.5 to 100 GHz},'' 3rd
  Generation Partnership Project, TR 38.901 V16.1.0, Dec. 2019.

\bibitem{wu20IRS}
Q.~{Wu} and R.~{Zhang}, ``{Towards Smart and Reconfigurable Environment:
  Intelligent Reflecting Surface Aided Wireless Network},'' \emph{IEEE
  Communications Magazine}, vol.~58, no.~1, pp. 106--112, Jan. 2020.

\bibitem{pometcu18EuCAP}
L.~{Pometcu} and R.~D. {Errico}, ``{Characterization of sub-THz and mmwave
  propagation channel for indoor scenarios},'' in \emph{12th European
  Conference on Antennas and Propagation (EuCAP 2018)}, April 2018, pp. 1--4.

\bibitem{jacob09EUCAP}
M.~{Jacob} and T.~{Kurner}, ``{Radio channel characteristics for broadband
  WLAN/WPAN applications between 67 and 110 GHz},'' in \emph{3rd European
  Conference on Antennas and Propagation}, March 2009, pp. 2663--2667.

\bibitem{guan19TVT}
K.~{Guan} \emph{et~al.}, ``{Channel Characterization for Intra-Wagon
  Communication at 60 and 300 GHz Bands},'' \emph{IEEE Transactions on
  Vehicular Technology}, vol.~68, no.~6, pp. 5193--5207, June 2019.

\bibitem{ma18channel}
J.~Ma, R.~Shrestha, L.~Moeller, and D.~M. Mittleman, ``Channel performance for
  indoor and outdoor terahertz wireless links,'' \emph{APL Photonics}, vol.~3,
  no.~5, pp. 1--13, Feb. 2018.

\bibitem{abbasi20ICC}
N.~A. {Abbasi}, A.~{Hariharan}, A.~M. {Nair}, A.~S. {Almaiman}, F.~B.
  {Rottenberg}, A.~E. {Willner}, and A.~F. {Molisch}, ``Double directional
  channel measurements for thz communications in an urban environment,'' in
  \emph{2020 IEEE International Conference on Communications}, June 2020, pp.
  1--6.

\bibitem{nguyen2018comparing}
S.~L.~H. Nguyen \emph{et~al.}, ``{Comparing Radio Propagation Channels Between
  28 and 140 GHz Bands in a Shopping Mall},'' \emph{European Conference on
  Antennas and Propagation}, pp. 1--5, Apr. 2018.

\bibitem{viswanathan20A}
H.~{Viswanathan} and P.~E. {Mogensen}, ``{Communications in the 6G Era},''
  \emph{IEEE Access}, vol.~8, pp. 57\,063--57\,074, March 2020.

\bibitem{ghosh195g}
A.~{Ghosh}, A.~{Maeder}, M.~{Baker}, and D.~{Chandramouli}, ``{5G Evolution: A
  View on 5G Cellular Technology Beyond 3GPP Release 15},'' \emph{IEEE Access},
  vol.~7, pp. 127\,639--127\,651, Sept. 2019.

\bibitem{Mac17JSACb}
G.~R. MacCartney and T.~S. Rappaport, ``A flexible millimeter-wave channel
  sounder with absolute timing,'' \emph{IEEE Journal on Selected Areas in
  Communications}, vol.~35, no.~6, pp. 1402--1418, June 2017.

\bibitem{xing18GC}
Y.~Xing and T.~S. Rappaport, ``{Propagation Measurement System and Approach at
  140 GHz-Moving to 6G and Above 100 GHz},'' in \emph{\textit{IEEE 2018 Global
  Communications Conference}}, Dec. 2018, pp. 1--6.

\bibitem{Mac15b}
G.~R. {MacCartney, Jr.} \emph{et~al.}, ``Indoor office wideband millimeter-wave
  propagation measurements and models at 28 {GHz} and 73 {GHz} for ultra-dense
  {5G} wireless networks ({Invited Paper}),'' \emph{IEEE Access}, vol.~3, pp.
  2388--2424, Oct. 2015.

\bibitem{Rap02a}
T.~S. Rappaport, \emph{Wireless Communications: Principles and Practice},
  2nd~ed.\hskip 1em plus 0.5em minus 0.4em\relax Upper Saddle River, NJ:
  Prentice Hall, 2002.

\bibitem{rappaport2015wideband}
T.~S. Rappaport \emph{et~al.}, ``Wideband millimeter-wave propagation
  measurements and channel models for future wireless communication system
  design (invited paper),'' \emph{IEEE Tran. Comm.}, vol.~63, no.~9, pp.
  3029--3056, Sept. 2015.

\bibitem{sun2015path}
S.~Sun \emph{et~al.}, ``Path loss, shadow fading, and line-of-sight probability
  models for {5G} urban macro-cellular scenarios,'' in \emph{2015 IEEE Globecom
  Workshops (GC Wkshps)}, Dec. 2015, pp. 1--7.

\bibitem{Rap15a}
T.~S. Rappaport \emph{et~al.}, \emph{Millimeter Wave Wireless
  Communications}.\hskip 1em plus 0.5em minus 0.4em\relax Pearson/Prentice
  Hall, 2015.

\bibitem{Samimi15b}
M.~K. Samimi \emph{et~al.}, ``Probabilistic omnidirectional path loss models
  for millimeter-wave outdoor communications,'' \emph{IEEE Wireless
  Communications Letters}, vol.~4, no.~4, pp. 357--360, Aug. 2015.

\bibitem{Sun14a}
S.~Sun \emph{et~al.}, ``Millimeter wave multi-beam antenna combining for {5G}
  cellular link improvement in {New York City},'' in \emph{2014 IEEE
  International Conference on Communications (ICC)}, June 2014, pp. 5468--5473.

\bibitem{ju19icc}
S.~Ju \emph{et~al.}, ``{Scattering Mechanisms and Modeling for Terahertz
  Wireless Communications},'' in \emph{Proc.\ IEEE International Conference on
  Communications (ICC)}, May 2019, pp. 1--7.

\bibitem{sun2015synthesizing}
S.~Sun \emph{et~al.}, ``{Synthesizing Omnidirectional antenna patterns,
  received power and path loss from directional antennas for 5G millimeter-wave
  communications},'' in \emph{IEEE GLOBECOM}, Dec. 2015, pp. 3948--3953.

\bibitem{samimi20163}
M.~K. Samimi and T.~S. Rappaport, ``{3-D millimeter-wave statistical channel
  model for {5G} wireless system design},'' \emph{IEEE Transactions on
  Microwave Theory and Techniques}, vol.~64, no.~7, pp. 2207--2225, July 2016.

\end{thebibliography}

\end{document}